%% file: asonam24.tex
\documentclass{llncs}
\usepackage[T1]{fontenc}

\usepackage{hyperref}

\usepackage{color}

\usepackage{tikz}\usetikzlibrary{positioning, graphs, patterns, graphs.standard}
\usepackage{subcaption}

\usepackage{amsfonts,amssymb}
\usepackage{mathtools}

\usepackage[noadjust]{cite}

\usepackage{algorithm}
\usepackage{algpseudocode}
\usepackage{bookmark}
\algdef{S}[FOR]{ForAll}[1]{\algorithmicforeall\ #1\ \algorithmicdo}

\begin{document}

\title{Methodology for Identifying Social Groups within a Transactional Graph}

\author{Maxence Morin\inst{1, 2}\orcidID{0009{-}0006{-}3789{-}9904} \and
Baptiste Hemery\inst{2}\orcidID{0000{-}0003{-}3923{-}1829} \and
Fabrice Jeanne\inst{2} \and
Estelle Pawlowski-Cherrier\inst{1}\orcidID{0009{-}0008{-}9872{-}3667}}

\authorrunning{M. Morin et al.}

\institute{Normandie Univ, UNICAEN, ENSICAEN, CNRS, GREYC, 14000 Caen, France \\
\email{maxence.morin@unicaen.fr} \\
\email{estelle.pawlowski@ensicaen.fr} \and
Orange Innovation, 14000 Caen, France\\
\email{\{baptiste.hemery,fabrice.jeanne\}@orange.com}}

\maketitle

\begin{abstract}
Social network analysis is pivotal for organizations aiming to leverage the vast amounts of data generated from user interactions on social media and other digital platforms.
These interactions often reveal complex social structures, such as tightly-knit groups based on common interests, which are crucial for enhancing service personalization or fraud detection.
Traditional methods like community detection and graph matching, while useful, often fall short of accurately identifying specific groups of users.
This paper introduces a novel framework specifically designed to identify groups of users within transactional graphs by focusing on the contextual and structural nuances that define these groups.
\end{abstract}

\section{Introduction}
Social network analysis is subject to increasing attention.
Relational data are obtained from the interactions of users of a service.
They are of great value.
Whether it comes from social media, telecommunications services, or financial transactions, companies in these sectors have understood the importance of extracting information from relational data.
They exploit information to enrich and personalize services, improve trust by combating fraud and improper use, and so on.
A common method to analyze relational data is to transform data into transactional graphs.
Nodes represent users, and edges represent transactions.
This abstraction allows us to solve some issues in social network analysis thanks to graph theory.

In these graphs, we search for groups of users that present interesting characteristics.
Users exchange in a manner that creates communities.
These can be identified using community detection algorithms.
However, in this paper we are not interested in community detection.
We are looking for other specific types of groups in transactional graphs.

We define a new concept that we call Group of Interest (GoI).
A GoI is a social group composed of users who share social relations.
The difference between communities and GoI is explained in the following paragraphs.
As with communities, GoI can overlap.
For example, a user can be a member of a family and a member of a group of fraudsters at the same time.
We define the term ``social truth'' to encompass all social groups present in transactional graphs.
In the real world, we cannot know the entirety of the social truth.
Interviewing people costs money.
Thus, we only know a part of the social truth that we call \(samples\).

We refer to all the exchanges observed in a service as the ``transactional truth''.
Transactions are traces of underlying social relationships.
However, transactional truth is bound by a period of time.
This restriction makes transactional truth incomplete, and it may reflect a part or the entirety of the social truth.
If we extend the period of time, every social relationship may induce a transaction.
In this paper, we aim to uncover the social truth by using only the transactional truth.

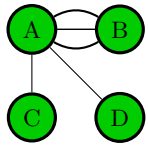
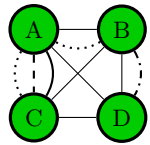
\begin{figure*}[htbp!]
\centering
\begin{subfigure}[t]{0.45\textwidth}
\centering
\input{fig_sgi_type_a_w_context.tex}
\caption{Observed exchanges in a transactional service}%
\label{fig:sgi_type_a_w_context}
\end{subfigure}
\begin{subfigure}[t]{0.45\textwidth}
\centering
\input{fig_sgi_type_a_reality.tex}
\caption{Underlying social relationships between the users.
Different types of social relationships are represented by different line styles.
}%
\label{fig:sgi_type_a_reality}
\end{subfigure}
\caption{Differences between the observed transactional truth and the underlying and inaccessible social truth.}%
\label{fig:reality_observed_truth}
\end{figure*}

One could naturally think of traditional community detection methods to solve this specific problem, but these methods are not sufficient for two main reasons.
The first reason is that these algorithms identify communities, which are groups of nodes that are highly interconnected relative to the rest of the graph.
The transactions induced by a GoI do not create a highly connected subgraph.
For example, in the subgraph shown in \autoref{fig:sgi_type_a_w_context}, each node \(A,B,C, \text{or}\, D\) does not engage in transactions with everyone.
We are far from having a densely connected subgraph.
Nonetheless, the users represented by these nodes may know each other and all share a relationship.
The underlying relationships may be dense, as illustrated in \autoref{fig:sgi_type_a_reality}.
The second reason is that community detection methods cover the entire graph and attempt to identify communities of any type, whereas we are interested in only specific subgraphs.

The second natural solution one might also think of could belong to the graph matching research domain.
Graph matching methods usually require a query to find topologically equivalent subgraphs.
But the query does not consider any context surrounding it.
We consider that the context is important to increase the performance of the model.
Nevertheless, community detection methods as well as graph matching methods can represent a part of our solution to social truth identification.
We give some details in \autoref{sec:subgraphs} to adapt these methods in a new and complete framework.

In this paper, we introduce the notion of a SubGraph of Interest (SGI).
Furthermore, we propose a new framework that encompasses the social truth identification problem and formalizes several methods for solving it.
We also propose a method to evaluate the framework's performance.
The experimental evaluation of the framework will be presented in a further paper.

The rest of this paper is organized as follows.
\autoref{sec:preliminaries} is devoted to the problem statement.
The framework is presented in \autoref{sec:framework_definition}.
\autoref{sec:evaluation} describes how to evaluate the framework.
Conclusions and further work are discussed in \autoref{sec:conclusions}.

\section{Problem Statement}\label{sec:preliminaries}

\subsection{Notations}\label{sec:notations}
Let \(G = (V_G,E_G)\) be a multigraph comprised of a set of nodes \(V_G\) and edges \(E_G\).
Each vertex \(v \in V_G\) represents a user.
Each edge \((u,v) \in E_G\) denotes a relationship such as a financial transaction, a communication, or a friendship.
Let \(S = (V_S, E_S)\) be a subgraph, which implies \(S \subseteq G\) if \(V_S \subseteq V_G\) and \(E_S \subseteq E_G\).

\subsection{Problem Formalization}\label{sec:problem_formalization}
We define a new concept that we call SubGraph of Interest (SGI).
An SGI is a subgraph that represents the transactions made by a GoI.
Let \({\cal{S}}_{type}\) be an SGI that represents transactions of a certain type of GoI.
Similar to GoIs, SGIs can overlap.
Thus, a node can belong to several SGIs.
Let \(\mathbb{S}_{type}\) be the set containing all the \({\cal{S}}_{type}\) of a unique type.
The set \(\mathbb{S}_{type}\) is representing one GoI type present in the social truth.
The type can be families, groups of colleagues, fraudsters, and so on.
For the sake of clarity, we use \(\mathbb{S}\) and \({\cal{S}}\) instead of \(\mathbb{S}_{type}\) and \({\cal{S}}_{type}\) as we are only interested in a unique type of GoI at a time.
Recovering the set \(\mathbb{S}\) for each type means, in the end, recovering the entire social truth.

Let \(\mathbb{S}_n \subset \mathbb{S}\) be the set of known \({\cal{S}}\) in \(\mathbb{S}\) that represents our \(samples\).
The set \(\mathbb{S}_n\) has a fixed size \(n\).
These \(n\) subgraphs are used as training samples.
As mentioned in the introduction, we have to deal with a set \(\mathbb{S}_n\) containing very few samples.

The aim of this paper is to recover the entire set \(\mathbb{S}\) using only the training sample set \(\mathbb{S}_n\) for one given type.
The contribution of this paper is twofold:
\begin{enumerate}
\item Propose a framework to tackle the previously mentioned problem of social truth identification.
\item Evaluate this framework.
\end{enumerate}

\section{Framework Definition}\label{sec:framework_definition}
Given a multigraph \(G\) and a set of samples \(\mathbb{S}_n\), we want the framework to output a set \(\hat{\mathbb{S}}\).
\(\hat{\cal{S}}\) stands for the elements of \(\hat{\mathbb{S}}\).
We want this set \(\hat{\mathbb{S}}\) to be as close as possible to the set \(\mathbb{S}\).

We illustrate the framework as follows:
\begin{equation}
(G,\mathbb{S}_n)
\to
\boxed{{method}}
\to \hat{\mathbb{S}}
\;\text{such that}\;
\hat{\mathbb{S}}
\simeq
\mathbb{S}
\label{eq:framework}
\end{equation}

The relation \(\hat{\mathbb{S}} \simeq \mathbb{S}\) means that we want the set \(\hat{\mathbb{S}}\) to be as similar as possible to the social truth \(\mathbb{S}\) (which is only part of the social truth, i.e.\ for a specific type).
We precise this point in \autoref{sec:evaluation}.

Now we define three \(method\) proposals in framework~\eqref{eq:framework}.
The first two \(methods\) share the same basis and are gathered in the first approach, detailed in \autoref{sec:first_approach}.
The last \(method\) relies on a new pruning approach, detailed in \autoref{sec:second_approach}.

\subsection{First Approach}\label{sec:first_approach}

The \(method\) of the first approach consists of three steps.

\begin{enumerate}
\item The first step is a function named \textsc{subgraphs}.
This function takes a multigraph \(G\) as input and must return a set of clusters.
The role of this function can be fulfilled by already existing functions.
We chose functions from either community detection or graph matching research domains.
\item The second step is the \textsc{features} function.
This function takes a subgraph as input and should return a fixed-size vector that encodes information about the given subgraph.
Details about this function are provided in a further paragraph.
\item The last step of the first approach is the selection step.
The clusters returned by the \textsc{subgraphs} function are not necessarily relevant.
We need to select those that are similar enough to the available \(samples\) in \(\mathbb{S}_n\).
\end{enumerate}

We present this \textsc{FirstApproach} function through the \autoref{algo:first_approach}.
We go through the details of each step now.

\begin{algorithm}
\input{algo_fa.tex}
\caption{\textsc{FirstApproach} and \textsc{check} functions.}%
\label{algo:first_approach}
\end{algorithm}

\subsubsection{Subgraphs Function}\label{sec:subgraphs}
The role of the \textsc{subgraphs} function is to create subgraphs from the given multigraph \(G\).
It seems natural to think of community detection and graph matching methods as good functions for creating clusters.

On the one hand, community detection algorithms~\cite{Fortunato2010} yield subgraphs, also referred to as communities.
They identify either overlapping or non-overlapping communities.
Because SGI can overlap, we are interested in overlapping community detection.

Among the overlapping community detection algorithms, we find algorithms using edge betweenness~\cite{Girvan2002} or modularity based algorithms~\cite{Singh2022}.
We propose to use an overlapping variant of the Label Propagation algorithm~\cite{Zhu2002,Liu2021} for its inherent usage of the neighborhood.
It creates communities by considering nodes interactions with their neighbors.
The algorithm uses a threshold to determine a node's membership in a community.
It also analyzes the frequency to set the dominant label to propagate.

Nevertheless, we identified two limitations to community detection as a solution for identifying the social truth.
The first limitation stems from the definition of a community itself.
Often, a community is referred to as a subgraph that is highly connected among its nodes and that shares only a few edges with the rest of the graph~\cite{Fortunato2010}.

The second limitation is the design of community detection algorithms.
These algorithms create clusters of any type, covering the entire graph and providing subgraphs even if they are not relevant.
In a sense, community detection algorithms provide more information than we require, as our problem is confined to a specific type of SGI.\@

On the other hand, graph matching~\cite{Yan2016} consists in finding all occurrences of a query subgraph within a graph.
An occurrence is a subgraph that has the same topology as the query.
The problem that consists of ensuring that a subgraph has the same topology as a query is named the isomorphism problem~\cite{Grohe2020}.
This problem is NP-complete in the general case~\cite{Babai2016}.

As well as for community detection based methods, considering the problem of social truth identification, one can point out three limitations to graph matching based methods.
The first limitation is that a query cannot represent an SGI in \(\mathbb{S}\).
A query has a fixed number of nodes and edges.
SGIs in \(\mathbb{S}\) have high variability in the number of nodes and edges.
To overcome this limitation, we can look for Approximate Graph Matching~\cite{Yan2016, Gallagher2006}.
Approximate Graph Matching methods find subgraphs that are similar enough to a query.
The similarity can be computed using the graph edit distance.
The graph edit distance is the required distortion to obtain the subgraph by inserting, removing, or substituting some nodes or edges.
However, this metric is generally NP-hard to compute~\cite{Zeng2009}.

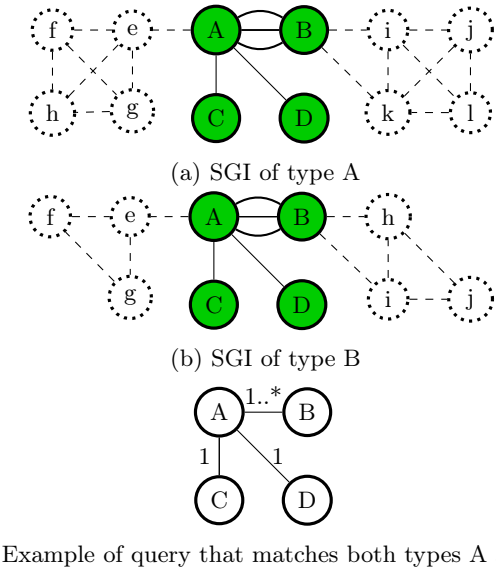
\begin{figure}[htbp]
\centering

\begin{subfigure}{\textwidth}
\centering
\input{fig_sgi_type_a.tex}
\caption{SGI of type A}%
\label{fig:sgi_type_a}
\end{subfigure}

\begin{subfigure}{\textwidth}
\centering
\input{fig_sgi_type_b.tex}
\caption{SGI of type B}%
\label{fig:sgi_type_b}
\end{subfigure}

\begin{subfigure}{\textwidth}
\centering
\input{query.tex}
\caption{Example of query that matches both types A and B.}%
\label{fig:query}
\end{subfigure}

\caption{
The first two subgraphs represent two different types of SGI.\@
The difference is not in the topology of the subgraphs but in their context.
A simple query like in the subgraph~\ref{fig:query} is not enough to differentiate them.
}%
\label{fig:sgi_types}
\end{figure}

The second limitation of this approach in our use case is the lack of context around the query subgraph.
In a social network, a group of people is defined by its structure, its personal pieces of information, and the way it connects to the rest of the graph.
For example, the query in \autoref{fig:query} matches both types A and B in \autoref{fig:sgi_types}.
The only way to distinguish them is by considering their extended neighbors.
For this reason, a query should be sampled from the graph to take the context around it into account.
To the best of our knowledge, such an approach does not exist in the literature.

The third limitation is that there are \(n\) samples in \(\mathbb{S}_n\), whereas graph matching requires a unique query.
We compensate for this limitation by analyzing subgraphs in \(\mathbb{S}_n\) and computing their Maximum Common Subgraph (MCS)~\cite{Roy2024} to obtain a unique query.
Then, we propose to do classic graph matching with this MCS as the query.
This approach is the one we chose for the \textsc{subgraphs} function when considering graph matching.

\subsubsection{Features Function}\label{sec:features}
A feature vector enables encoding information about a graph, a subgraph, a node, or an edge.
This concept is related to element characterization in graphs.
The objective of characterization is to generate a vector of fixed size that encodes information about an element~\cite{Li2012,Perozzi2014,Alsentzer2020,Grover2016,Narayanan2017}.
In the \textsc{FirstApproach} function presented in \autoref{algo:first_approach}, the \textsc{features} function takes a subgraph as input.
Thus, we are interested in subgraph characterization.

Subgraph characterization requires capturing the local structure, the subgraph's relative position, and nodes' specific information.
We achieve this by defining a vector composed of some metrics, as mentioned in reference~\cite{Li2012}.
For example, we can use the following metrics:
\begin{itemize}
\item The number of nodes and edges in the subgraph.
These metrics are straightforward and discriminate between subgraphs of different sizes.
\item The minimum, maximum, and average of degrees, or clustering coefficients encode the internal structure of the graph.
\item The minimum, maximum, and average of the shortest path length within the subgraph encode the diameter of the subgraph and can detect central nodes.
\item Transitivity: This is the fraction of all possible triangles present in the subgraph.
It encodes how tightly knitted the subgraph is and gives information about the subgraph's internal structure.
\item Assortativity is the similarity of connections in the graph with respect to the node degree.
This metric informs us how the connected nodes are similar to each other.
\end{itemize}

This list is non-exhaustive.
The selection of one metric over another is related to the dataset and the SGI type we are searching for.

In the state of the art, we find deep learning-based methods to create feature vectors.
We find methods such as SubGNN~\cite{Alsentzer2020}, SUGAR~\cite{Sun2021}, or Sub2Vec~\cite{Adhikari2018}.

\subsubsection{Selection Step}\label{sec:selection}
The \textsc{subgraphs} function returns subgraphs that might be irrelevant.
For this reason, we have to select only those that could match a subgraph present in \(\mathbb{S}\).
We propose to resort to the cosine distance \(d_{c}(A, B) = \frac{A \cdot B}{\lVert A \rVert\lVert B \rVert}\) and define an upper bound \(\Gamma\) for the distance to make sure that our subgraphs are relevant.
This guarantees that the property \(\hat{\mathbb{S}} \simeq \mathbb{S}\) in framework~\eqref{eq:framework} is fulfilled.
This distance is applied to the feature vectors previously defined.
Other distances, such as the Euclidean distance or the Manhattan distance, could also be considered.

\subsection{Second Approach}\label{sec:second_approach}
In this section, we propose a \(method\) with a completely different strategy.
Instead of predicting where the subgraphs are and whether elements such as nodes or edges belong to an SGI or not, we approach the problem in reverse.
We are interested in predicting elements that are not in an SGI.\@
We then prune these elements from the graph.
We suppose that the resulting connected components are all SGIs in \(\mathbb{S}\).
This approach is presented in \autoref{fig:second_approach}.

\begin{algorithm}[htb!]
\input{algo_sa.tex}
\caption{\textsc{SecondApproach} function.}%
\label{fig:second_approach}
\end{algorithm}

We can relate this approach to the edge betweenness community detection algorithm~\cite{Girvan2002}.
The difference is that our approach is not limited to edge pruning.
To prune elements, we first need to know which elements do not belong to an SGI in \(\mathbb{S}\).
This is achieved by comparing the features of elements of the multigraph with elements from samples in \(\mathbb{S}_n\).
Again, we need feature vectors for nodes and edges in the graph.
These vectors can be obtained using two methods.

The first method involves using a simple hand-engineered feature vector, as presented in \autoref{sec:features}, applied at both the node and edge levels.
We design feature vectors using users' and transactions' properties.
The second method uses graph embedding techniques such as the Graph Isomorphism Network (GIN)~\cite{Xu2019} for node features and the Edge2Vec method~\cite{Wang2020} for edge features.

GIN is based on a Message Passing Neural Network~\cite{Kipf2017}.
It creates a feature vector based on the neighborhood of nodes.
GIN is as powerful as the Weisfeiler-Lehman graph isomorphism test~\cite{Weisfeiler1968}, which ensures that the topology around nodes is correctly encoded.

Edge2vec combines a deep autoencoder~\cite{Salakhutdinov2009} to reduce the size of the feature vector and a Skipgram model~\cite{Mikolov2013} to ensure the vector correctly encodes the context around edges.

We compare the features of nodes and edges of the multigraph with the training samples in \(\mathbb{S}_n\).
We propose using again the cosine distance \(d_c\) as defined in \autoref{sec:selection}.
If a node or an edge differs too much from any of the training samples, we add it to the \(bad\) set.
Let \({\cal{F}}\) represent the features of an element, such as a node or an edge.
We consider thresholds \(\Gamma_{node}\) and \(\Gamma_{edge}\) as the maximum distance a node or an edge can be from a node or an edge of the training samples.
The sets \(V_{bad}\) and \(E_{bad}\), referred to as the \(bad\) sets, contain the elements to be removed from the multigraph \(G\).
We consider all the connected components obtained by the function \textsc{connected} to be the results \(\hat{\mathbb{S}}\).

We define a pruning function \(\nabla\) that takes a multigraph \(G\), the bad sets \(V_{bad}\) and \(E_{bad}\), and returns a graph from which the nodes and edges from \(V_{bad}\) and \(E_{bad}\) have been removed.
We express the \(\nabla\) function as follows:
\begin{equation}
\nabla(G, V_{bad}, E_{bad}) = G_{pruned}
\end{equation}
We define four specific pruning functions: \(\nabla_{simple}\), \(\nabla_{node}\), \(\nabla_{edge}\), and \(\nabla_{majority}\).

The \(\nabla_{simple}\) pruning function removes all the nodes and edges present in the \(bad\) sets.
We express the \(\nabla_{simple}\) pruning function as follows:
\begin{equation}
\nabla_{simple}(G, V_{bad}, E_{bad}) = (V_G \setminus V_{bad}, E_G \setminus E_{bad})
\end{equation}
When we remove a node, we also remove all edges that were connected to that node.

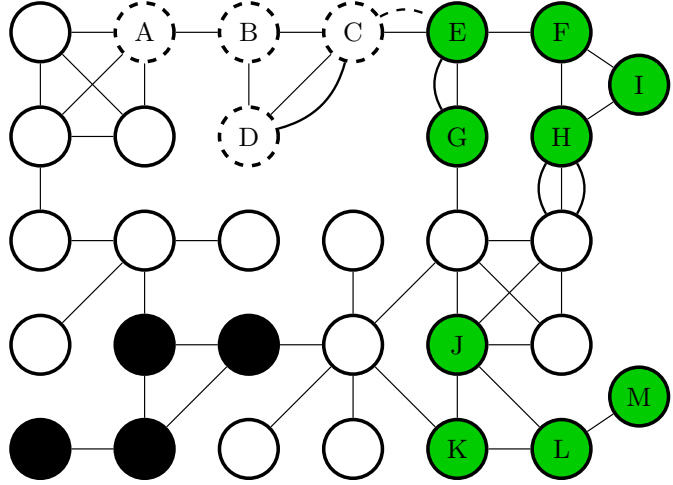
\begin{figure}[htbp!]
\centering
\resizebox{0.75\linewidth}{!}{
\input{fig.tex}
}
\caption{
An example of a multigraph.
Dashed nodes are from a subgraph in \(\mathbb{S}_n\).
Dashed and green nodes compose three subgraphs from \(\mathbb{S}\).
Black nodes are from a subgraph that shares a similar topology but not the same context.
}%
\label{fig:example}
\end{figure}

The \(\nabla_{node}\) is expressed as follows:
\begin{equation}
\nabla_{node}(G, V_{bad}, E_{bad}) = (V_G \setminus V_{bad}, E_G)
\end{equation}
The weakness of the \(\nabla_{node}\) pruning function is the case of two connected subgraphs.
For example, in \autoref{fig:example}, we cannot distinguish the subgraph \(ABCD\) from the subgraph \(EFGHI\) by removing nodes only.
We need to remove the edge \((C, E)\) to reveal the two connected components.

We formalize the \(\nabla_{edge}\) pruning function as follows:
\begin{equation}
\nabla_{edge}(G, V_{bad}, E_{bad}) = (V_G, E_G \setminus E_{bad})
\end{equation}
The weakness of the \(\nabla_{edge}\) pruning function is the multigraph context.
We need to remove all edges between two nodes to create two disconnected components.
For example, in \autoref{fig:example}, in the case of multiple edges between nodes \(C\) and \(E\) represented by a dashed line, all edges between \(C\) and \(E\) must be removed.

The \(\nabla_{majority}\) pruning function removes nodes and edges, but spares nodes if the majority of their edges are not in \(E_{bad}\).
We define the function as follows:
\begin{equation}
edgeMajority(v) = \frac{|E(v) \setminus E_{bad}|}{|E(v)|}
\end{equation}
Here, \(E(v)\) represents the set of all edges connected to node v, and \(E(v) \setminus E_{bad}\) represents the set of edges connected to \(v\) that are not in \(E_{bad}\).
The ratio thus gives the proportion of good edges to total edges connected to \(v\).
Next, we define the set of all spared nodes as follows:
\begin{equation}
V_{spared} = \{v \in V_G \;|\; edgeMajority(v) \geq 0.5\}
\end{equation}
In this expression, \(V_G\) represents the set of all nodes in the multigraph.
The condition \(edgeMajority(v) \geq 0.5\) ensures that only those nodes for which the majority of their edges are not in \(E_{bad}\) are included in \(V_{spared}\).
Finally, we can formalize the pruning function as follows:
\begin{equation}
\nabla_{majority}(G, V_{bad}, E_{bad}) = (V_G \setminus (V_{bad} \setminus V_{spared})\,,\, E_G \setminus E_{bad})
\end{equation}

The expression \((V_{bad} \setminus V_{spared})\) identifies the set of nodes that are bad and not spared.
Essentially, it removes the spared nodes from the set of bad nodes, ensuring that nodes with a majority of good connections are not pruned.

This \(method\) of pruning nodes and edges from the multigraph is a new perspective on the problem of social truth identification.
We reversed the problem of the SGIs position's prediction by focusing on the removal of elements from the multigraph to create connected components.

\section{How to Evaluate the Framework}\label{sec:evaluation}
In this section, we propose an adaptation of the \(precision\) and \(recall\) metrics to evaluate the performance of the \(method\) in the framework~\eqref{eq:framework}.
We need to compare the output \(\hat{\mathbb{S}}\) of the framework with the social truth \(\mathbb{S}\).
We describe a \(match\) function that is used in the adaptation of the \(precision\) and the \(recall\).
The \(match\) function is a boolean function.
It guarantees that two subgraphs are similar enough to be matched.
It takes \(\hat{\cal{S}} \in \hat{\mathbb{S}}\) and \({\cal{S}} \in \mathbb{S}\) as inputs and returns \(true\) if they match, \(false\) otherwise.
The \(match\) function uses the three boolean functions \(\varphi_{extra}\), \(\varphi_{missing}\), and \(\varphi_{size}\) that we describe now.

\begin{enumerate}
\item The \(\varphi_{extra}\) boolean function checks that the ratio of the nodes found in \(\hat{\cal{S}}\) that are not in \({\cal{S}}\) over the size of \(V_{\cal S}\) is under the threshold \(\Gamma_{extra}\).
We express it as follows:
\begin{equation}
    \varphi_{extra}(\hat{\cal S}, {\cal S}) = \frac{|V_{\hat{\cal{S}}} \setminus V_{\cal S}|}{|V_{\cal S}|} < \Gamma_{extra}
\end{equation}
\item The \(\varphi_{missing}\) boolean function checks that the ratio of nodes in \({\cal{S}}\) that are not present in \(\hat{{\cal{S}}}\) over the number of nodes in \(V_{\cal{S}}\) is under the threshold \(\Gamma_{missing}\).
We express it as follows:
\begin{equation}
    \varphi_{missing}(\hat{\cal S}, {\cal S}) = \frac{|V_{\cal{S}} \setminus V_{\hat{\cal{S}}} |}{|V_{\cal{S}}|} < \Gamma_{missing}
\end{equation}
\item The \(\varphi_{size}\) boolean function checks that the size difference between \({\cal{S}}\) and \(\hat{\cal{S}}\) over the size of \(V_{\cal S}\) is under a threshold \(\Gamma_{size}\).
We express it as follows:
\begin{equation}
    \varphi_{size}(\hat{\cal S}, {\cal S}) = \frac{abs(|V_{\hat{\cal{S}}}| - |V_{\cal{S}}| )}{|V_{\cal{S}}|} < \Gamma_{size}
\end{equation}
\end{enumerate}
Every threshold is set according to the context of the problem.

We need to consider the three boolean functions together to declare two subgraphs as matching.
We express the \(match\) function as follows:
\begin{equation}
match({\cal{S}},\hat{\cal{S}}) =
\left\{
\begin{matrix}
\varphi_{extra}(\hat{\cal S}, {\cal S})   & \And \\
\varphi_{missing}(\hat{\cal S}, {\cal S}) & \And \\
\varphi_{size}(\hat{\cal S}, {\cal S})
\end{matrix}
\right.
\end{equation}

Iverson bracket~\cite{Iverson1962} is a notation that generalizes the Kronecker delta.
This function is defined to take the value 1 if the statement is true and the value 0 otherwise.
It is generally denoted as follows, where \(P\) is the statement:
\begin{equation}
[P] =
\left\{
\begin{matrix}
1 & \text{if} \;P\; \text{is true}; \\
0 & \text{otherwise}.
\end{matrix}
\right.
\end{equation}

We define a subgraph \({\cal{S}}_a\) as relevant if it matches at least one subgraph in a set of subgraphs \(\Psi\).
The \(\Psi\) variable can take the value \(\mathbb{S}_n\), \(\mathbb{S}\), or \(\hat{\mathbb{S}}\).
We use the Iverson bracket to formalize the \(relevant\) boolean function as follows:
\begin{equation}
relevant({\cal{S}}_a,\Psi) = \big(\sum_{{\cal{S}}_b \in \Psi} [match({\cal{S}}_a,{\cal{S}}_b)]\big) \geq 1
\end{equation}

We use \(precision\) and \(recall\) metrics to evaluate the framework.
The \(precision\) metric counts the proportion of relevant SGIs among selected SGIs.
The \(recall\) metric counts the proportion of relevant SGIs selected among all selectable relevant SGIs.
We define it as the ratio of \(\hat{\cal{S}}\) that matches an element in \(\mathbb{S}\) out of all the predicted elements in \(\hat{\mathbb{S}}\).
We formalize \(precision\) as follows:
\begin{equation}
precision = \frac{\sum_{\hat{\cal{S}} \in \hat{\mathbb{S}}}[relevant(\hat{\cal{S}}, \mathbb{S})] }{|\hat{\mathbb{S}}|}
\end{equation}
A higher precision means that the function finds more subgraphs that are close enough to the social truth.

We formalize \(recall\) as follows:
\begin{equation}
recall = \frac{\sum_{{\cal{S}}\in\mathbb{S}}[relevant({\cal{S}},\hat{\mathbb{S}})] }{|\mathbb{S}|}
\end{equation}
A higher recall means that a greater number of the social truth subgraphs are matched.

In conclusion, the adaptation of precision and recall metrics, integrated with the match function, provides a robust framework for evaluating the performance of the method in the framework~\eqref{eq:framework}.
We can also calculate an F-measure score~\cite{Hand2021} such as an F1-score~\cite{Sokolova2006}.

\section{Conclusions and Future Work}\label{sec:conclusions}
In this paper, we introduce a novel framework designed to address the challenge of social truth identification within transactional graphs.
This is a critical task for deciphering complex social interactions in large datasets.
We present three distinct methods that combine traditional graph analysis techniques with innovative approaches to recover the hidden social truth.
By adapting precision and recall metrics, we have established a method to evaluate the performance of the proposed framework.
This ensures that the identified SubGraphs of Interest are relevant and representative of the underlying social structures.

For the future, we propose two interesting directions that involve integrating more advanced deep learning techniques into our framework:
\begin{enumerate}
\item Enhancement of Community Detection: Incorporating deep learning techniques into community detection algorithms could significantly improve their performance across diverse datasets.
Exploring methods such as the Neural Overlapping Community Detection model~\cite{Shchur2019} and other deep learning-based approaches could offer substantial improvements in detecting structures within multigraphs.
\item Deep Learning as a distance function: Applying deep learning to enhance the distance function \(d\) used with hand-crafted feature vectors and a pruning approach could increase the framework's robustness and performance.
This integration would potentially enable more accurate identification of various types of SGIs, making the framework more effective across different scenarios.
\end{enumerate}

Experiments using the framework are in progress and will be presented in a separate article.
We will present the performance and efficiency in terms of time and memory of the framework with large datasets.

\begin{credits}
\subsubsection{\ackname} The authors would like to thank Orange and the ANRT for funding the thesis.
\end{credits}

\bibliographystyle{splncs04}
\bibliography{bibi}
\end{document}

%% file: fig_sgi_type_a_w_context.tex
\begin{tikzpicture}[
        cnode/.style={circle, fill=black!20!green, draw = black, very thick},
        edge/.style={draw,thick,-}
    ]

    \node[cnode] (A) {A};
    \node[cnode] (B) [below=0.5cm of A] {C};
    \node[cnode] (C) [right=0.5cm of A] {B};
    \node[cnode] (D) [below=0.5cm of C] {D};

    \draw (A) -- (B);
    \path[edge] (A) to[bend right] (C);
    \path[edge] (A) to[bend left] (C);
    \draw (A) -- (C);
    \draw (A) -- (C);
    \draw (A) -- (D);
    \draw (A) -- (C);
    \draw (A) -- (C);

\end{tikzpicture}

%% file: fig_sgi_type_a_reality.tex
\begin{tikzpicture}[
    cnode/.style={circle, fill=black!20!green, draw = black, very thick},
    edge/.style={draw,thick,-}
]

\node[cnode] (A) {A};
\node[cnode] (B) [right=0.5cm of A] {B};
\node[cnode] (C) [below=0.5cm of A] {C};
\node[cnode] (D) [below=0.5cm of B] {D};

\draw (A) -- (B);
\path[edge, dotted] (A) to[bend right] (C);
\path[edge] (A) to[bend left] (C);
\path[edge, dashed] (A) to (C);

\path[edge, dashdotted] (B) to[bend left] (D);
\path[edge, dotted] (B) to[bend left] (A);

\draw (A) -- (D);

\draw (C) -- (D);
\draw (B) -- (D);
\draw (B) -- (C);

\end{tikzpicture}

%% file: algo_fa.tex
\begin{algorithmic}[1]
\algdef{S}[FOR]{ForAll}[1]{\algorithmicforall\ #1\ \algorithmicdo}
\Function{FirstApproach}{G, $\mathbb{S}_n$, \textsc{subgraphs}, $\Gamma$}
\State$\hat{\mathbb{S}} \gets $ \Call{subgraphs}{G}
\State$results \gets $ \Call{List}{$\empty$}
\ForAll{$\hat{\cal{S}} \in \hat{\mathbb{S}} $}
\State$\hat{{\cal{F}}} \gets$\Call{features}{$\hat{{\cal{S}}}$}

\If{\Call{check}{$\hat{{\cal{F}}},\mathbb{S}_n,\Gamma$}}
\State\(results \gets results + \hat{\cal{S}}\)
\State\textbf{break}
\EndIf%
\EndFor%
\State\textbf{return} \(results\)
\EndFunction%

\Function{check}{${\cal{F}}_\kappa,\Psi,\Gamma$}
\ForAll{\(\psi \in \Psi\)}
\State${\cal{F}}_\psi \gets$\Call{features}{$\psi$}
\If{$d_c({\cal{F}}_\kappa, {\cal{F}}_\psi) < \Gamma$}
\State\textbf{return} \(true\)
\EndIf%
\EndFor%
\State\textbf{return} \(false\)
\EndFunction%
\end{algorithmic}

%% file: fig_sgi_type_a.tex
\begin{tikzpicture}[
    cnode/.style={circle, fill=black!20!green, draw = black, very thick},
    edge/.style={draw,thick,-}
]

\node[cnode] (A) {A};
\node[cnode] (B) [below=0.5cm of A] {C};
\node[cnode] (C) [right=0.5cm of A] {B};
\node[cnode] (D) [below=0.5cm of C] {D};
\node[cnode] (E) [left=0.5cm of A, fill=white,dotted, minimum size=0.5cm] {e};
\node[cnode] (F) [left=0.5cm of E, fill=white,dotted, minimum size=0.5cm] {f};
\node[cnode] (G) [below=0.5cm of E,fill=white,dotted, minimum size=0.5cm] {g};
\node[cnode] (H) [below=0.5cm of F,fill=white,dotted, minimum size=0.5cm] {h};

\node[cnode] (I) [right=0.5cm of C,fill=white,dotted, minimum size=0.5cm] {i};
\node[cnode] (J) [right=0.5cm of I,fill=white,dotted, minimum size=0.5cm] {j};
\node[cnode] (K) [below=0.5cm of I,fill=white,dotted, minimum size=0.5cm] {k};
\node[cnode] (L) [below=0.5cm of J,fill=white,dotted, minimum size=0.5cm] {l};

\draw (A) -- (B);
\path[edge] (A) to[bend right] (C);
\path[edge] (A) to[bend left] (C);
\draw (A) -- (C);
\draw (A) -- (C);
\draw (A) -- (D);
\draw (A) -- (C);
\draw (A) -- (C);
\draw[dashed] (A) -- (E);
\draw[dashed] (E) -- (G);
\draw[dashed] (G) -- (H);
\draw[dashed] (F) -- (E);
\draw[dashed] (F) -- (H);

\draw[dashed] (C) -- (I);
\draw[dashed] (I) -- (J);
\draw[dashed] (I) -- (K);
\draw[dashed] (L) -- (K);
\draw[dashed] (C) -- (K);
\draw[dashed] (L) -- (J);

\draw[dashed] (F) -- (G);
\draw[dashed] (H) -- (E);

\draw[dashed] (I) -- (L);
\draw[dashed] (J) -- (K);

\end{tikzpicture}

%% file: fig_sgi_type_b.tex
\begin{tikzpicture}[
        cnode/.style={circle, fill=black!20!green, draw = black, very thick},
        edge/.style={draw,thick,-}
    ]
    \node[cnode] (A) [below=0.5cm of B]{A};
    \node[cnode] (B) [below=0.5cm of A] {C};
    \node[cnode] (C) [right=0.5cm of A] {B};
    \node[cnode] (D) [below=0.5cm of C] {D};

    \node[cnode] (E) [left=0.5cm of A,fill=white,dotted, minimum size=0.5cm] {e};
    \node[cnode] (F) [left=0.5cm of E,fill=white,dotted, minimum size=0.5cm] {f};
    \node[cnode] (G) [below=0.5cm of E,fill=white,dotted, minimum size=0.5cm] {g};

    \node[cnode] (H) [right=0.5cm of C,fill=white,dotted, minimum size=0.5cm] {h};
    \node[cnode] (I) [below=0.5cm of H,fill=white,dotted, minimum size=0.5cm] {i};
    \node[cnode] (J) [right=0.5cm of I,fill=white,dotted, minimum size=0.5cm] {j};

    \draw (A) -- (B);
    \path[edge] (A) to[bend right] (C);
    \path[edge] (A) to[bend left] (C);
    \draw (A) -- (C);
    \draw (A) -- (C);
    \draw (A) -- (D);
    \draw (A) -- (C);
    \draw (A) -- (C);
    \draw[dashed] (A) -- (E);
    \draw[dashed] (E) -- (G);
    \draw[dashed] (G) -- (F);
    \draw[dashed] (F) -- (E);

    \draw[dashed] (C) -- (H);
    \draw[dashed] (C) -- (I);
    \draw[dashed] (I) -- (J);
    \draw[dashed] (H) -- (J);
    \draw[dashed] (I) -- (H);
\end{tikzpicture}

%% file: query.tex
\begin{tikzpicture}[
        cnode/.style={circle, fill=white, draw = black, very thick},
        edge/.style={draw,thick,-}
    ]

    \node[cnode] (A) {A};
    \node[cnode] (B) [right=0.5cm of A] {B};
    \node[cnode] (C) [below=0.5cm of A] {C};
    \node[cnode] (D) [below=0.5cm of B] {D};

    \draw (A) -- (B) node[midway,above] {1..*};
    \draw (A) -- (C) node[midway,left] {1};
    \draw (A) -- (D)  node[midway,right] {1};
    \draw (A) -- (C);

\end{tikzpicture}

%% file: algo_sa.tex
\begin{algorithmic}[1]
    \algdef{S}[FOR]{ForAll}[1]{\algorithmicforall\ #1\ \algorithmicdo}
    \Function{SecondApproach}{G, $\mathbb{S}_n$, $\Gamma_{node}, \Gamma_{edge}$}
    \State $V_{bad} \gets V_G$
    \State $E_{bad} \gets E_G$
    \ForAll {$u \in V_G$}
    \State ${\cal{F}}_u \gets$\Call{features}{$u$}
    \ForAll {${\cal{S}} \in \mathbb{S}_n$}
    \If{\Call{check}{${\cal{F}}_u,V_{\cal{S}},\Gamma_{node}$}}
    \State $V_{bad} \gets V_{bad} - u$
    \State \textbf{break}
    \EndIf
    \EndFor
    \EndFor

    \ForAll {$e \in E_G$}
    \State ${\cal{F}}_e \gets$\Call{features}{$e$}
    \ForAll {${\cal{S}} \in \mathbb{S}_n$}
    \If{\Call{check}{${\cal{F}}_e,E_{\cal{S}},\Gamma_{edge}$}}
    \State $E_{bad} \gets E_{bad} - e$
    \State \textbf{break}
    \EndIf
    \EndFor
    \EndFor
    \State $G_{pruned} \gets$ \Call{$\nabla$}{$G,V_{bad},E_{bad}$}
    \State \textbf{return} \Call{connected}{$G$}
    \EndFunction
\end{algorithmic}

%% file: fig.tex
\begin{tikzpicture}[
        cnode/.style={circle, fill=white, draw = black, very thick, minimum size=0.7cm},
        cblue/.style={fill=blue!50},
        cgreen/.style={fill=black!20!green},
        cpink/.style={fill=white, dashed},
        corange/.style={fill=orange},
        cred/.style={fill=black},
        legendnode/.style={shape=circle, draw=black, line width=1},
        edge/.style={draw,thick,-}
    ]

    \node[cnode]                                          (1) {};
    \node[cnode, right=0.5cm of 1, cpink]                 (2) {A};
    \node[cnode, below=0.5cm of 1]                        (3) {};
    \node[cnode, below=0.5cm of 2]                        (4) {};
    \node[cnode, right=0.5cm of 2, cpink]                 (5) {B};
    \node[cnode, right=0.5cm of 5, cpink]                 (6) {C};
    \node[cnode, below=0.5cm of 5, cpink]                 (7) {D};

    \node[cnode, right=0.5cm of 6, cgreen]                (8) {E};
    \node[cnode, right=0.5cm of 8, cgreen]                (9) {F};
    \node[cnode, below right=0.1 and 0.4 of 9, cgreen]    (10) {I};
    \node[cnode, below=0.5cm of 8, cgreen]                (11) {G};
    \node[cnode, below=0.5cm of 9, cgreen]                (12) {H};

    \node[cnode, below=0.5cm of 11]                       (13) {};
    \node[cnode, below=0.5cm of 12]                       (14) {};
    \node[cnode, below=0.5cm of 14]                       (15) {};
    \node[cnode, below=0.5cm of 13, cgreen]               (16) {J};
    \node[cnode, below=0.5cm of 16, cgreen]               (17) {K};
    \node[cnode, right=0.5cm of 17, cgreen]               (18) {L};
    \node[cnode, above right=0.1 and 0.4 of 18, cgreen]   (19) {M};
    \node[cnode, left=0.5cm of 17]                        (20) {};
    \node[cnode, left=0.5cm of 20]                        (21) {};
    \node[cnode, above=0.5cm of 21, cred]                 (22) {};
    \node[cnode, left=0.5cm of 16]                        (23) {};
    \node[cnode, above=0.5cm of 23]                       (24) {};

    \node[cnode, left=0.5cm of 22, cred]                  (25) {};
    \node[cnode, above=0.5cm of 22]                       (26) {};
    \node[cnode, above=0.5cm of 25]                       (27) {};
    \node[cnode, left=0.5cm of 27]                        (28) {};
    \node[cnode, below=0.5cm of 28]                       (29) {};
    \node[cnode, below=0.5cm of 29, cred]                 (30) {};
    \node[cnode, right=0.5cm of 30, cred]                 (31) {};

    \draw (1) -- (2);
    \draw (1) -- (3);
    \draw (1) -- (4);
    \draw (2) -- (3);
    \draw (2) -- (4);
    \draw (3) -- (4);
    \draw (2) -- (5);
    \draw (5) -- (6);
    \draw (5) -- (7);
    \draw (6) -- (7);
    \path[edge] (6) to[bend left] (7);
    \draw (6) -- (8);
    \path[edge, dashed] (6) to[bend left] (8);

    \draw (8) -- (9);
    \draw (8) -- (11);
    \path[edge] (8) to[bend right] (11);
    \draw (9) -- (10);
    \draw (9) -- (12);
    \draw (10) -- (12);
    \draw (11) -- (13);
    \draw (12) -- (14);
    \path[edge] (12) to[bend right] (14);
    \path[edge] (12) to[bend left] (14);
    \draw (13) -- (14);
    \draw (13) -- (15);
    \draw (13) -- (16);
    \draw (14) -- (15);
    \draw (14) -- (16);
    \draw (15) -- (16);
    \draw (16) -- (17);
    \draw (16) -- (18);
    \draw (17) -- (18);
    \draw (18) -- (19);
    \draw (17) -- (23);
    \draw (20) -- (23);
    \draw (21) -- (23);
    \draw (22) -- (23);
    \draw (23) -- (24);

    \draw (3)  -- (28);
    \draw (22) -- (31);
    \draw (22) -- (25);
    \draw (25) -- (31);
    \draw (25) -- (27);
    \draw (26) -- (27);
    \draw (27) -- (28);
    \draw (27) -- (29);
    \draw (30) -- (31);
    \draw (13) -- (23);

\end{tikzpicture}